# Optimum design for BB84 quantum key distribution in tree-type passive optical networks

José Capmany[1,*] and Carlos R. Fernández-Pousa[2]

[1] *ITEAM Research Institute, Universidad Politécnica de Valencia, 46022 Valencia, Spain*

[2] *Signal Theory and Communications, Dep. of Physics and Computer Science, Univ. Miguel Hernández, 03202 Elche, Spain*

[*]*Corresponding author: jcapmany@iteam.upv.es*

We show that there is a tradeoff between the useful key distribution bit rate and the total length of deployed fiber in tree-type passive optical networks for BB84 quantum key distribution applications. A two stage splitting architecture where one splitting is carried in the central office and a second in the outside plant and figure of merit to account for the tradeoff are proposed. We find that there is an optimum solution for the splitting ratios of both stages in the case of Photon Number Splitting (PNS) attacks and Decoy State transmission. We then analyze the effects of the different relevant physical parameters of the PON on the optimum solution.

# 1 Introduction

The objective of Quantum Key Distribution (QKD) is to provide an unique way of sharing a random sequence of bits between users with a level of security not attainable with either public or secret-key classical cryptographic systems [1, 2]. In essence, QKD relies on exploiting the laws of quantum mechanics [3, 4]. Most of the reported experimental results on long distance QKD rely on different photonic-based techniques and are based on the so-called BB84 protocol. For instance, in 1992 Bennett and co-workers [5] proposed to exploit the polarization of photons to implement the four required states by employing one circular polarization and one linear polarization basis. Later, Townsend and co-workers [6-8], proposed the use of optical delays and balanced interferometers at the transmitter and the receiver. A third approach, based on differential phase shift quantum key distribution [9] has enabled key generation and distribution along distances over 100 km [10] although with limited security [11]. Finally, a fourth approach [12], also known as frequency coding, relies on encoding the information bits on the sidebands of either phase [13] or amplitude [14] radio-frequency (RF) modulated light.

Much of the work reported in the literature has been focused towards point to point key distribution but, as pointed in [7, 15], to find truly widespread application QKD techniques should be employed in communication networks where any-to-any and any-to many transmission can occur. In particular, a first scenario where this may happen is in fiber based passive optical networks, where the passive nature (no optical amplification) and the limited distance range (up to 20 km) favors the implementation of multiuser BB84 QKD systems. In this context, recent contributions [16] have addressed the comparison of different multiuser quantum key distribution schemes over PONs, paying especial emphasis on the attainable quantum bit error rate $Q$ but not addressing the issue of combined $Q$ or even more important, the final secure key *R* that can be extracted from the sifted key [17] and the optimization of resources which, even in the most simple PON configurations, cannot be considered as uncoupled factors. Indeed the importance of this subject has been recently raised in an exhaustive review on the subject [18] in which it has been pointed out that in practical QKD "physical" figures of merit, such as secret key or maximal achievable distance are in competition with "practical" figures of merit, such as stability and cost.

To illustrate this point with an example, the upper part of figure 1 shows a typical N-user tree-PON configuration where Alice, at the central office is connected via an optical fiber link of length $L_1$ to a $1 \times N$ passive splitter. Each of the *N* outputs of the splitter connects to a different end-user (Bob$_i$, *i=1, 2…N*) via an optical fiber link of length $L_2$ (we will assume that the length of any Alice-Bob$_i$ connection is fixed and equal to $L=L_1+L_2$). With the exception of Japan, where the very high and homogeneous population density has dictated a point to point architecture from the central office, the typical Fiber to the Home (FTTH) access network scenario is formed by end users in close geographical proximity which are grouped into clusters. Each cluster is served by a local star coupler to which each user is connected by a short-length individual fiber [18]. As discussed elsewhere [15], the quantum level behavior of the splitter enables the key distribution task between Alice and the different Bobs, since a single photon incident on the splitter cannot be divided, but it will be randomly (and unpredictably) routed to one (and only one) of the output paths, with a probability given by *1/N*. If we assume that Alice and the different end-users employ a BB84 protocol for key distribution, based on any of the above reported photonic techniques, then it is easy to compute the end-to-end power transmission factor of a particular Alice-Bob$_i$ connection, which is given by $T_L = e^{-\alpha L}/N$, where α is the fiber attenuation constant. Also, we can have an estimation of the optical resources (i.e network cost) employed by calculating the total length of deployed fiber in the PON, which is given by $L_T = L_1 + NL_2$. To improve the power transmission factor one could think, for instance, in including the 1×N splitter inside the central office as shown in the lower part of figure 1. In this case, the power transmission factor is increased by a factor of *N*, that is, $T_L = e^{-\alpha L}$, but the total length of deployed fiber in the PON is increased up to $L_T = NL_1 + NL_2 = NL$. Thus, as it can be appreciated, increasing the end-to-end transmission factor (and thus decreasing its *Q* value and, in consequence, the final secure key *R*) of a given Alice-Bob$_i$ connection comes at the price of requiring more fiber to be deployed in the outside plant. It turns out that in the context of access networks the fiber installation costs have a very significant impact (up to an 85%) on the total deployment costs [18], thus, there is a design tradeoff between *R* and deployed fiber length in the tree-type PON, for which we wish to find an optimum solution. The purpose of this paper is to find such a solution and discuss the effect of the different and relevant physical parameters of the BB84 QKD system on it.

## 2 Tree PON Architecture Description

Figure 2 shows the proposed two splitting stage PON layout that is an intermediate case between the two previously discussed. Here, a first *1×N$_1$* passive splitter is located inside the central office at the output of Alice's transmitter and thus it does not contribute to the PON's loss. A different fiber link of length *L$_1$* connects each of the outputs from the *1×N$_1$* splitter to an *1×N$_2$* secondary splitter which, in turn, is connected by different fiber links of length *L$_2$* to *N$_2$* different final users (Bob$_i$, *i* =1, 2… *N*). We assume that Alice's transmitter and Bob$_i$'s (*i* = 1, 2,…*N*) are the required for the particular encoding method employed to implement the BB84 protocol (polarizations, phase or frequency).

Referring to figure 2 the end to end transmission from Alice to a particular Bob$_i$ due to system's losses is given by:

$$T_L = \frac{e^{-\alpha(L_1+L_2)}}{N_2} = \frac{T_F}{N_2} = \frac{N_1 T_F}{N} \tag{1}$$

while the total length of fiber employed to connect the final users to Alice is given by:

$$L_T = N_1 L_1 + N L_2 \tag{2}$$

Note that $N_1$ and $N_2$ are linked by the relationship $N=N_1 \times N_2$, where $N$ is the total number of end users. Increasing the value of $N_1$ (and thus decreasing the value of $N_2$) results in lower end-to end transmission losses but an increase in the total length of deployed fiber in the PON. Conversely, increasing the value of $N_2$ (and thus decreasing the value of $N_1$) results in higher end-to end transmission losses and a decrease in the total length of deployed fiber in the PON.

The values given by (1) and (2) lie in between those of the two extreme cases that we considered in the introduction section which optimize, respectively, the total length of fiber deployed (the most important factor in the network cost) and the end-to-end loss (i.e the final secure key $R$) . Since there is a trade-off between both parameters it is our aim to find an optimum configuration that can balance both contributions.

## 3 QKD Network Figure of Merit and Optimization

For optimization of the PON configuration it would be desirable to define a figure of merit which could take into account both physical and practical aspects as suggested in [18]. From the previous discussion a suitable magnitude fulfilling this criterion could be one considering both the effects of transmission losses and the total length of fiber deployed. Since the first directly impacts the QKD performance via the quantum bit error rate $Q$ and therefore the final secure key $R$ we propose the following expression:

$$FOM = \frac{R}{L_T} \tag{3}$$

For the computation of the final secure key rate we consider a system subject to photon Number Splitting (PNS) attack and decoy state transmission. We assume that there is a dominant decoy state whose average photon number is $\mu$, whereas the other decoy states are used to probe the channel. Then, the final secure key rate $R$ is related to the quantum bit error rate by [17,18]:

$$R \approx R_s \left[ e^{-\mu}(1-h(Q)) - h(Q) \right] \tag{4}$$

where $h(x) = -x\log_2(x) - (1-x)\log_2(1-x)$ is the binary entropy function, and $R_s = f_s \mu T_F \eta / N_2$ is the sifted key rate ($f_s$ is the pulse repetition rate, $T_F/N_2$ is the total loss, including the splitting loss, and $\eta$ is the detector efficiency). Here we assume that the value of $R_s$ is fixed, as any change in the total loss induced by different network plans could be compensated for, at low cost, by an adequate choice of repetition rate as long as it is below a certain value (10 GHz) limited by currently available off the shelf commercial devices. The quantity to be optimized per unit installed fiber's length is therefore the term in brackets in (4), which is a measure of the postprocessing (error correction and privacy amplification) required to extract an unconditionally secure key from the sifted key.

From (4) it follows that an increase in $Q$ results in a decrease of $R$ and viceversa. The reader can check that (3) is consistent with the trade-off previously described, since an increase in the total length of the deployed fiber reduces the *FOM* through the inverse dependence with $L_T$ but, at the same time, since end to end transmission losses are decreased ($N_2$ decreases) then so does the value of *Q*, (and therefore *R* increases), hence, the *R* factor tends to increase the FOM. On the other hand, a decrease in the total length of deployed fiber increases the FOM value through the ($1/L_T$) factor, while the increase in $N_2$ decreases the FOM through the *R* factor.

We can further develop the expression of the FOM showing its explicit dependence on $N_1$ by substituting (2), (4) into (3) and using the standard *Q* expression for BB84 systems which can be found elsewhere [3, 16]:

$$FOM = \frac{e^{-\mu}\left[1-h(Q)\right]-h(Q)}{N_1 L_1 + N L_2}$$

$$Q(N_1) = \frac{\mu \eta N_1 T_F (1-V) + N d_B}{2\mu \eta N_1 T_F} = \frac{1}{2}(1-V) + \frac{d_B}{2\mu \eta T_F}\frac{N}{N_1} \quad (5)$$

In the above expression *V* represents the optical visibility achieved in the filtering process, $\eta$ the detector efficiency, and $d_B$ is the dark count rate.

We can now optimize the FOM with respect to the branching ratio $N_1$ in the central office splitter:

$$\frac{\partial FOM}{\partial N_1} = 0 \Rightarrow$$
$$\Rightarrow \frac{N d_B (N_1 L_1 + N L_2)(1+e^{-\mu})}{2\mu \eta T_F} Ln(Q(N_1)) + N_1^2 L_1 \left[e^{-\mu} Ln(2) + (1+e^{-\mu})(Q(N_1) Ln(Q(N_1)) - Q(N_1))\right] = 0 \quad (6)$$

Eq.(6) is an implicit equation in $N_1$ that must be solved numerically and from which we can analyze the role that the different parameters play on the network design.

## 4 Results and Discussions

The effects of the different physical parameters on the optimum design of the branching ratios of the tree-PON architecture can be now investigated with the help of Eq. (6). Unless stated otherwise, we have taken the following typical values for $\lambda$=1550 nm: $d_B$ =$10^{-5}$, $V$ = 0.98, $\eta$=0.1, $L$ =$L_1$+$L_2$ = 20 km, and $\alpha$=0.25 dB/km. A caveat is pertinent at this point since although the link distance (20 km) may seem short in terms of losses, it should be taken into account that we are considering PONs with typical division ratios of *1×16, 1×32, 1×128*. In terms of insertion losses, for a given input-output connection, the former ratios are equivalent to adding extra links of 60, 75 and 105 km respectively at $\lambda$=1550 nm. This implies that the total transmission factor $T = T_F N_1/N$ corresponding to each point-to-point link Alice-Bob$_i$ lies in the range $T = 10^{-1.6} - 10^{-2.5}$ or 16-25 dB.

In order to estimate the value of the average photon number $\mu$ under which the PON is to be operated, we analyze the two contributions (5) to the QBER as follows. First, the visibility provides a constant value which, with the standard values used here, gives $Q_V$ =1%. This would be the only contribution to the QBER if the link is operated far from the threshold value where the net secure key rate drops to zero. If this would be the case, the optimal value of the average number of photons per pulse [18] is given by $\mu_{opt}$= 0.5 × [1−2h($Q_V$)] / [1−h($Q_V$)] = 0.46, which is slightly below the zero-QBER limit of 0.5. In practice, as we have already pointed out, the large equivalent point-to-point link loss in certain PON implementations precludes the use of these values without further justification.

The second term in (5) will be denoted $Q_D$ and accounts for the contribution of dark counts. It is straightforward to check that, in the worst-case scenario with $T$ = 25 dB the additional QBER due to dark counts is $Q_D$ < 3% only for $\mu$ > 0.5. In this case the total QBER is 4%, still far from the 11% limit of unconditional security. But the aforementioned optimal value of $\mu$ for transmission with decoy states gives $\mu_{opt}= 0.34$, a contradiction which simply shows that we have already reached the transmission threshold where the net security key rate drops to zero. However, for a slighter lower value of total loss of $T = 23$ dB we already obtain a consistent bound $\mu > 0.34$ for $Q_D < 3\%$. Assuming therefore that each Alice-Bob$_i$ link is operated with QBER in the range < 4% we have chosen a value $\mu=0.40$, which is representative of the range comprised by $\mu_{opt}= 0.50$ (zero-QBER limit), $\mu_{opt}= 0.46$ (visibility-dominated QBER) and $\mu_{opt}= 0.34$ (optimal for 4% QBER) and is also consistent with losses < 23 dB.

In addition, eq. (6) itself does not render integer values. Since optical splitters provide an integer number of output ports the results given by Eq. (6) have to be approximated taking into account the type of division provided by the splitters. Although *1×3, 1×5, 1×7* splitters are commercially available, we have chosen for our simulations splitters of the type *1×2$^I$*, where *I*=0,1,2,3... since these are the most commonly available in the market. Figures 3.a-3.d show the evolution of the figure of merit versus the value of $\log_2(N_1)$ given by (6) for the case where $L_1=15$ *km* and $L_2=5$ *km* and different values of the overall network branching number *N* (the values of the remaining parameters are those previously given). Note that, for each value of *N*, we obtain a value of $N_1$ ($N_1^{opt}$) that renders an optimum value of the figure of merit. It can be observed that the value of $N_1^{opt}$ increases with the value of *N*. This is because for given *N* the structure of (5) depends solely on ratio $N_1/N= 1/N_2$ and, for given *L*, on ratio $L_2/L_1$.

For a fixed value of *N* and *L* the value of $N_1^{opt}$ depends on the value of $L_1$. This is shown in figure 4 where we plot the value of $N_1^{opt}$ versus $L_1$ for different values of *N*. A similar behaviour is observed in all the cases with a decreasing behaviour as $L_1$ is increased.

Figure 5 shows as well the computed values for $Q$ (upper) and $R$ (lower) versus $L_1$, for the case where the number of users is 64. Note that for this particular configuration despite the maximum $Q$ values are comfortably below the 11% limit, reaching a maximum of around 3%, the value of $R/R_s$ can considerably decrease up to a 18%, thus it is this performance quantity and not the $Q$ which has to be considered when addressing the design of the network. For the sake of comparison and evaluation of the effect that the dark count rates have on the system performance, we have also included in figure 5 the same results obtained when considering an order of magnitude less, that is $d_B = 10^{-6}$ in the dark counts. The optimum system performance is much less sensitive to losses, as expected, since the second term in the QBER expression given by (5) which includes the effect of $N_1$ is one order of magnitude less significant. Furthermore, since the errors due to dark counts are less significant, the value of $R/R_s$ is higher.

Figure 6 shows the evolution of the optimum value of $\log_2(N_1)$ (upper) and $R/R_s$ (lower) in terms of $L_1$ for different values of $\mu$. For a given value of $L_1$ Eq. (6) yields an increase in the value of $N_1$ for decreasing values of $\mu$. This is due to the fact that when $\mu$ increases the end-to-end transmission factor increases (i.e. the QBER decreases or alternatively, $R$ increases). Thus, the same end-to-end performance can be achieved with a lower value of $N_1$. The variations in the value of $R$ increase with $L_1$ with more prominent step changes as the value of mean number of photons $\mu$ decreases.

## 5 Summary and Conclusions

In summary, we have shown that there is a tradeoff between useful key distribution bit rate $R$ and the total length of deployed fiber in tree-type passive optical networks for BB84 quantum key distribution applications. We have proposed a two stage splitting architecture where one splitting is carried in the central office and therefore does not contribute to system's loss and a second in the outside plant. We have proposed and justified a figure of merit to account for the tradeoff and found that there is an optimum solution in the case of Photon Number Splitting (PNS) attacks and Decoy State transmission for the splitting ratios of both stages. We have then analysed the effects of the different relevant physical parameters of the PON in the optimum solution.

## Acknowledgements

The authors wish to acknowledge the financial support of the Spanish Government through Quantum Optical Information Technology (QOIT), a CONSOLIDER-INGENIO 2010 Project and the Generalitat Valenciana through the PROMETEO research excellency award programme GVA PROMETEO 2008/092.

**Figure Captions**

**Figure 1.** Two tree-PON configurations for BB84-QKD. In the upper configuration the 1xN splitter affects the end-to end power transmission (and the QBER) between Alice and Bobi but minimum fiber resources need to be deployed. In the lower configuration the 1XN splitter does not affect the QBER but maximum fiber resources need to be deployed .

**Figure 2.** Proposed two-splitting-stage PON.

**Figure 3.** Evolution of the Figure of Merit in terms of $log_2(N_1)$, for a PON serving: 16 users (3.a), 32 users (3.b), 64 users (3.c) and 128 users (3.d).

**Figure 4.** Evolution of the optimum value of $log_2(N_1)$ in terms of $L_1$, for a PON serving16,32,64 and 128 users.

**Figure 5.** Computed values for $Q$ (upper) and $R$ (lower) versus $L_1$, for the case where the number of users is 64. Curves are shown for $d_B = 10^{-5}$ and $d_B = 10^{-4}$

**Figure 6.** Evolution of the optimum value of $log2(N_1)$ (upper) and $R$ (lower) in terms of $L_1$, for different values of $\mu$. The number of users is 64.

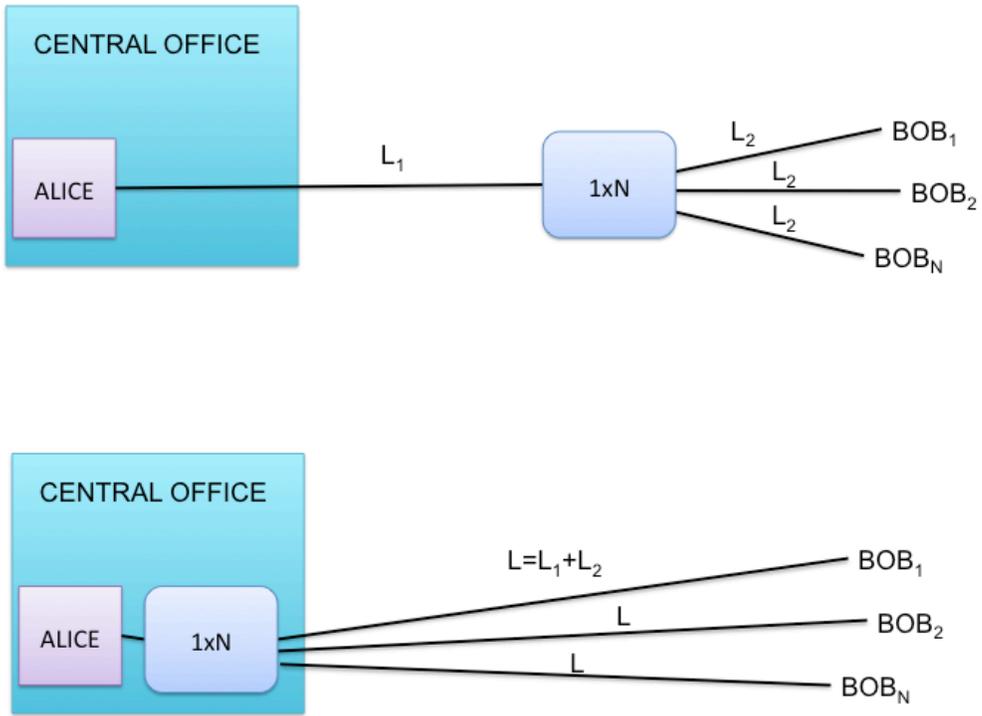

**FIGURE 1**

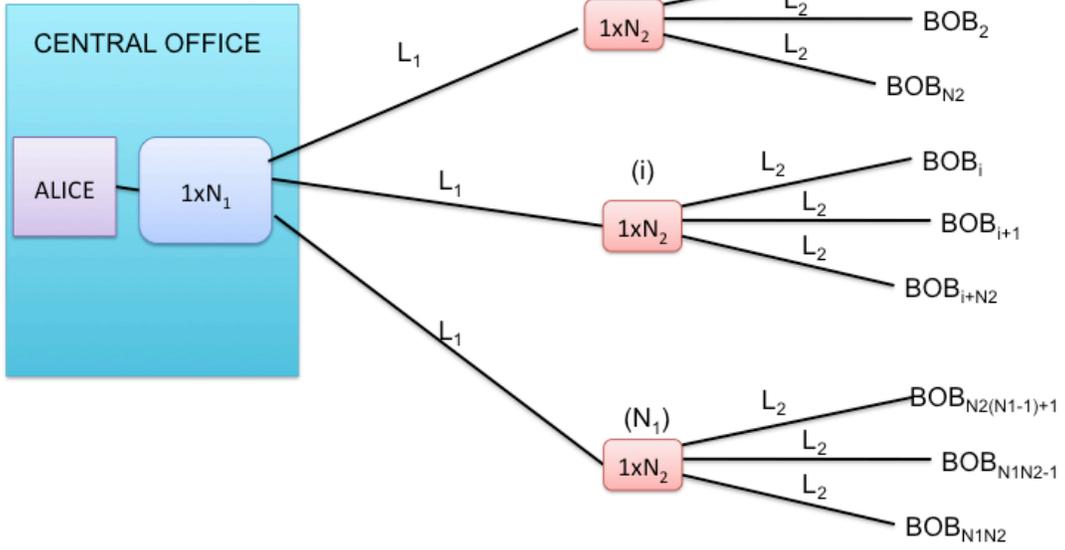

**FIGURE 2**

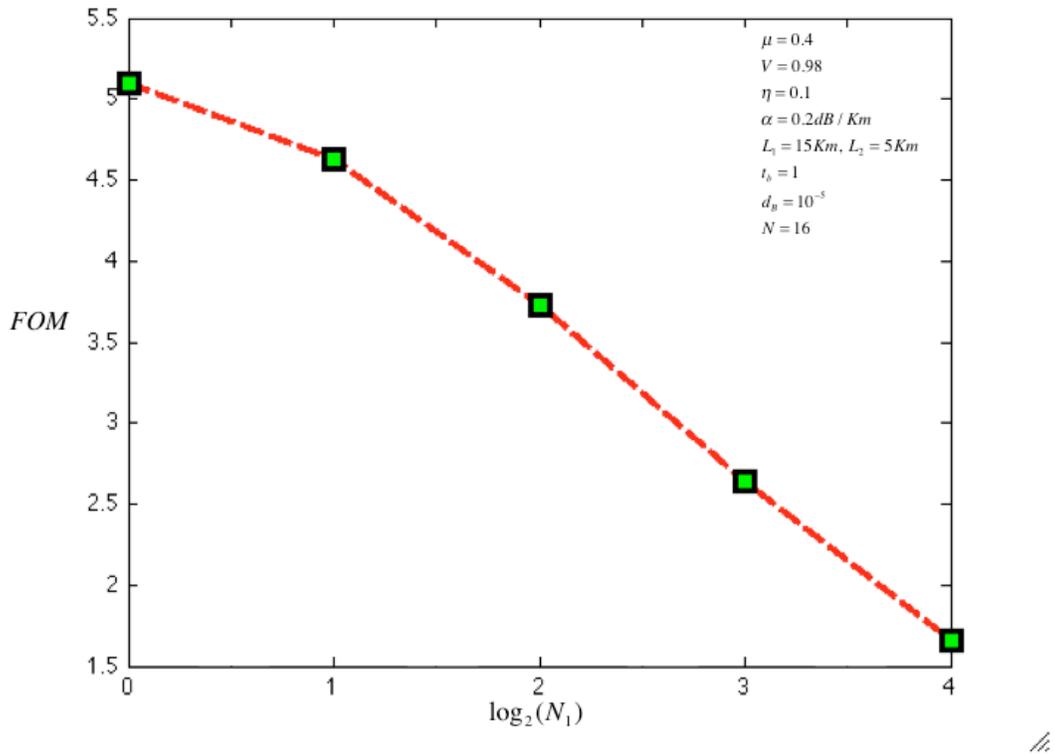

**FIGURE 3.a**

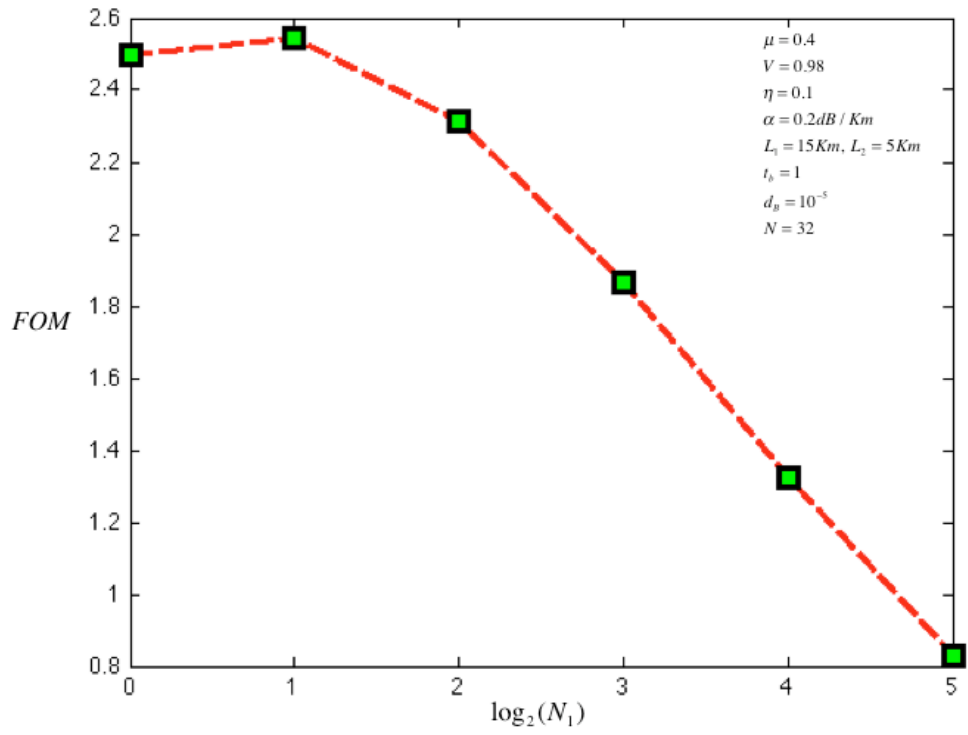

**FIGURE 3.b**

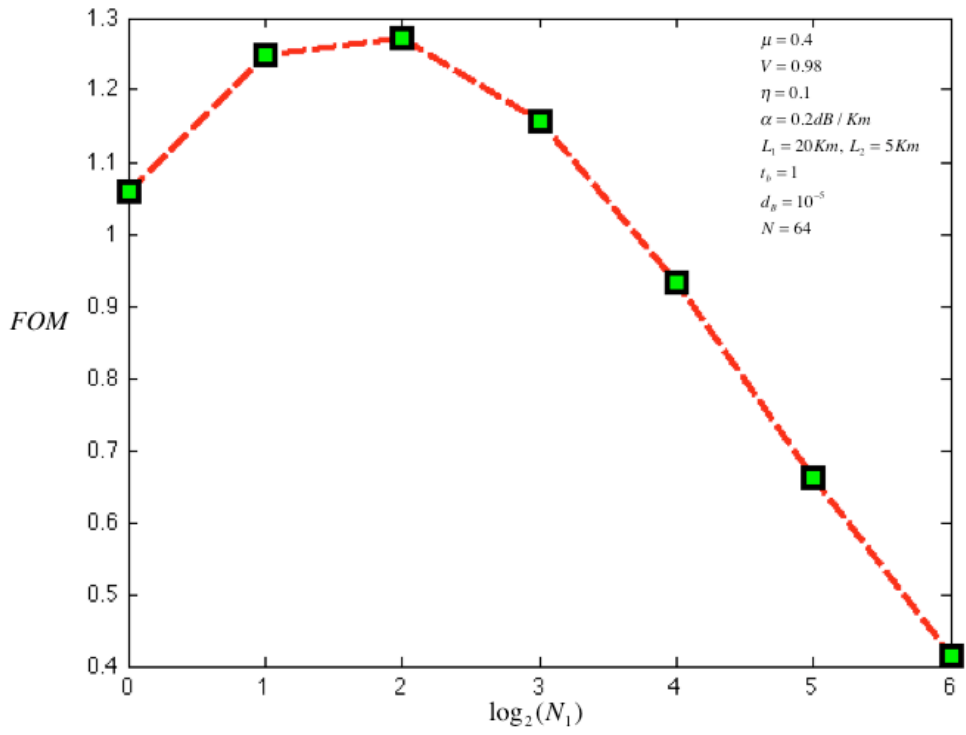

**FIGURE 3.c**

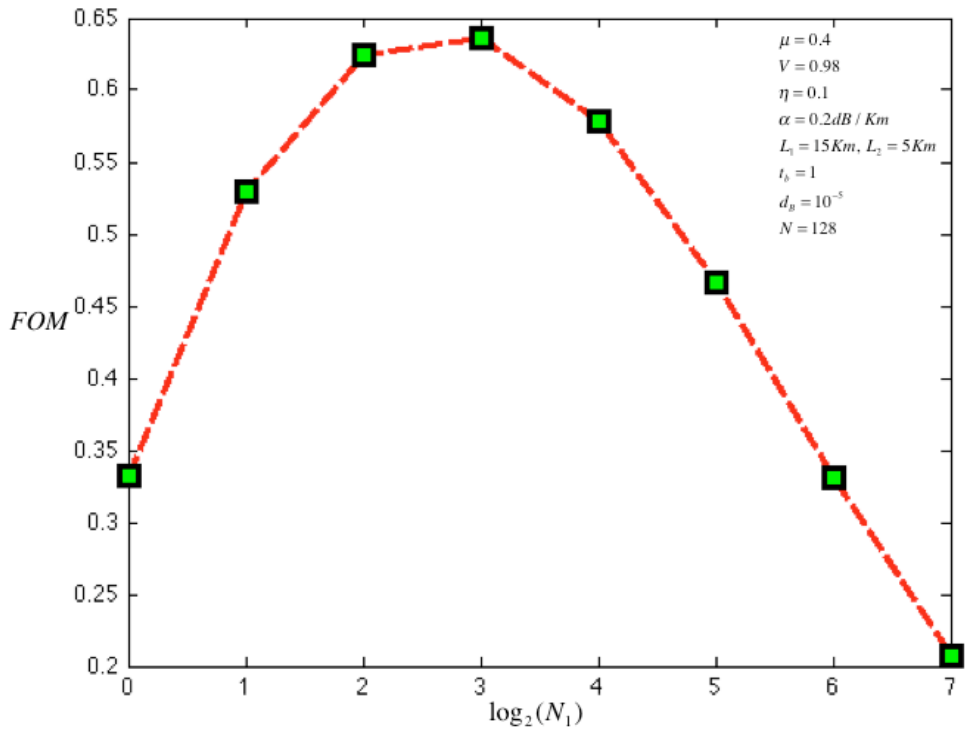

**FIGURE 3.d**

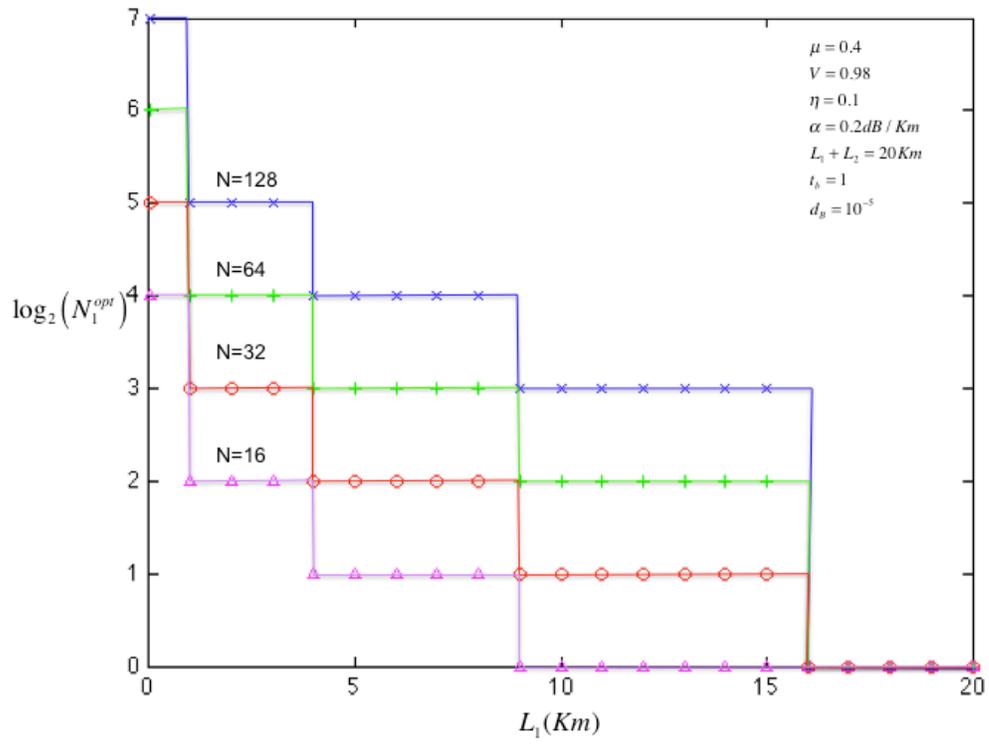

**FIGURE 4**

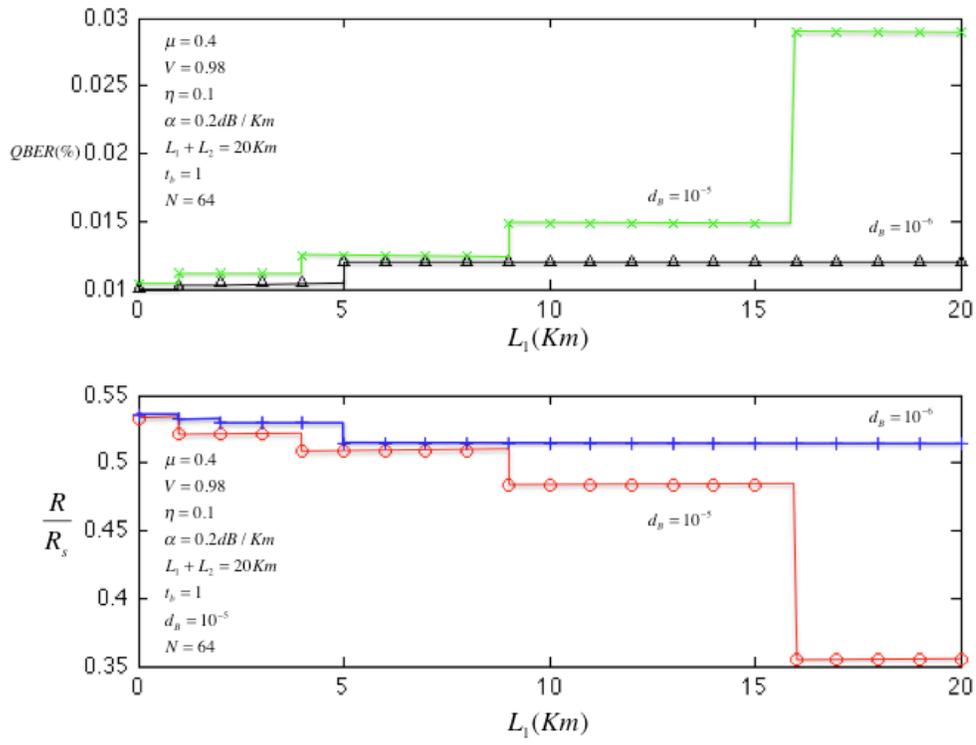

**FIGURE 5**

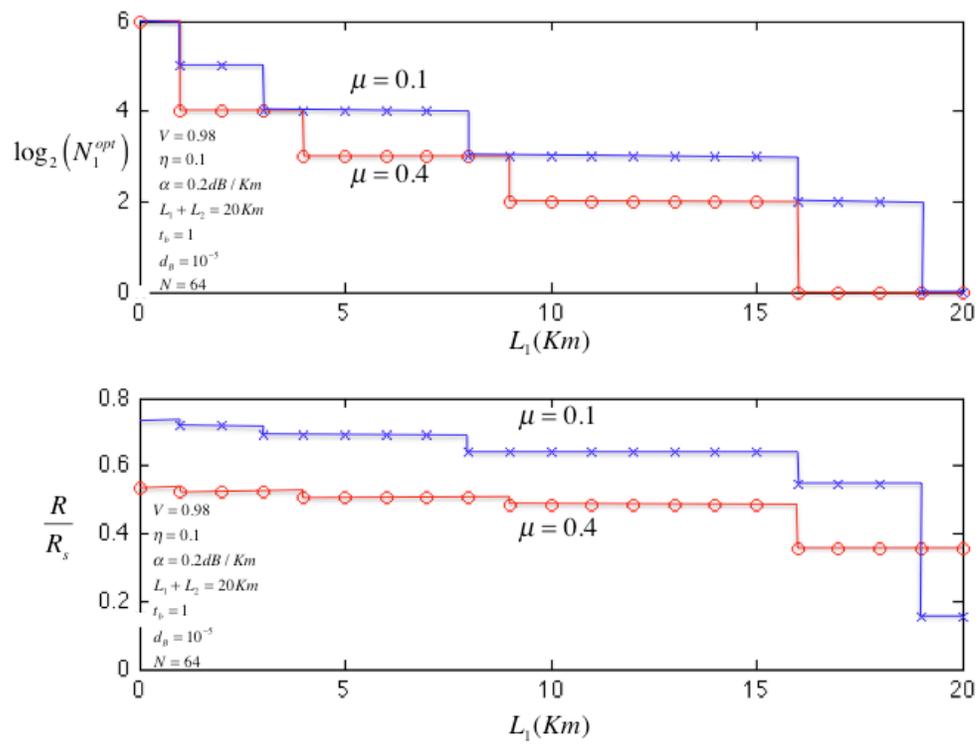

**FIGURE 6**